\documentclass[twoside,11pt,reqno]{amsart}
\usepackage{a4wide}
\usepackage{amsmath}
\usepackage{amsfonts}
\usepackage{amssymb}
\usepackage{times}
\usepackage[arrow, matrix, curve]{xy}
\usepackage{pstricks,pst-node,pst-text,pst-3d}
\usepackage{graphicx}
\catcode`\@11\relax
\newif\ify@autoscale \y@autoscaletrue \def\Yautoscale#1{\ifnum #1=0
  \y@autoscalefalse\else\y@autoscaletrue\fi}
\newdimen\y@b@xdim
\newdimen\y@boxdim \y@boxdim=13pt
\def\Yboxdim#1{\y@autoscalefalse\y@boxdim=#1}
\newdimen\y@linethick    \y@linethick=.3pt
\def\Ylinethick#1{\y@linethick=#1}
\newskip\y@interspace \y@interspace=0ex plus 0.3ex
\def\Yinterspace#1{\y@interspace=#1}
\newif\ify@vcenter   \y@vcenterfalse
\def\Yvcentermath#1{\ifnum #1=0 \y@vcenterfalse\else\y@vcentertrue\fi}
\newif\ify@stdtext   \y@stdtextfalse
\def\Ystdtext#1{\ifnum #1=0 \y@stdtextfalse\else\y@stdtexttrue\fi}
\newif\ify@enable@skew   \y@enable@skewfalse
\expandafter\ifx\csname enableskew\endcsname\relax
 \y@enable@skewfalse \else \y@enable@skewtrue\fi
\def\y@vr{\vrule height0.8\y@b@xdim width\y@linethick depth 0.2\y@b@xdim}
\def\y@emptybox{\y@vr\hbox to \y@b@xdim{\hfil}}
\ify@enable@skew
 \def\y@abcbox#1{\if :#1\else
   \y@vr\hbox to \y@b@xdim{\hfil#1\hfil}\fi}
 \def\y@mathabcbox#1{\if :#1\else
   \y@vr\hbox to \y@b@xdim{\hfil$#1$\hfil}\fi}
\else
 \def\y@abcbox#1{\y@vr\hbox to \y@b@xdim{\hfil#1\hfil}}
 \def\y@mathabcbox#1{\y@vr\hbox to \y@b@xdim{\hfil$#1$\hfil}}
\fi
\def\y@setdim{%
  \ify@autoscale%
   \ifvoid1\else\typeout{Package youngtab: box1 not free! Expect an
     error!}\fi%
   \setbox1=\hbox{A}\y@b@xdim=1.6\ht1 \setbox1=\hbox{}\box1%
  \else\y@b@xdim=\y@boxdim \advance\y@b@xdim by -2\y@linethick
  \fi}
\newcount\y@counter
\newif\ify@islastarg
\def\y@lastargtest#1,#2 {\if\space #2 \y@islastargtrue
  \else\y@islastargfalse\fi}
\def\y@emptyboxes#1{\y@counter=#1\loop\ifnum\y@counter>0
  \advance\y@counter by -1 \y@emptybox\repeat}
\def\y@nelineemptyboxes#1{%
  \vbox{%
    \hrule height\y@linethick%
    \hbox{\y@emptyboxes{#1}\y@vr}
    \hrule height\y@linethick}\vskip-\y@linethick}
\def\yng(#1){%
  \y@setdim%
  \hskip\y@interspace%
  \ifmmode\ify@vcenter\vcenter\fi\fi{%
  \y@lastargtest#1,
  \vbox{\offinterlineskip
    \ify@islastarg
     \y@nelineemptyboxes{#1}
    \else
     \y@ungempty(#1)
    \fi}}\hskip\y@interspace}
\def\y@ungempty(#1,#2){%
  \y@nelineemptyboxes{#1}
  \y@lastargtest#2,
  \ify@islastarg
   \y@nelineemptyboxes{#2}
  \else
   \y@ungempty(#2)
  \fi}
\def\y@nelettertest#1#2. {\if\space #2 \y@islastargtrue
  \else\y@islastargfalse\fi}
\def\y@abcboxes#1#2.{%
  \ify@stdtext\y@abcbox#1\else\y@mathabcbox#1\fi%
  \y@nelettertest #2.
  \ify@islastarg\unskip%
   \ify@stdtext\y@abcbox{#2}\else\y@mathabcbox{#2}\fi%
  \else\y@abcboxes#2.\fi}
 \newdimen\y@full@b@xdim
 \newcount\y@m@veright@cnt
\ify@enable@skew
 \def\y@get@m@veright@cnt#1#2.{%
   \if :#1 \advance\y@m@veright@cnt by 1\y@get@m@veright@cnt#2.\fi}
 \let\y@setdim@=\y@setdim
 \def\y@setdim{%
   \y@setdim@ \y@full@b@xdim=\y@b@xdim
   \advance\y@full@b@xdim by 1\y@linethick}
 \def\y@m@veright@ifskew#1{
   \y@m@veright@cnt=0 \y@get@m@veright@cnt#1.
   \moveright \y@m@veright@cnt\y@full@b@xdim}
\else
 \def\y@m@veright@ifskew#1{}
\fi
\def\y@nelineabcboxes#1{%
  \y@nelettertest #1.
  \ify@islastarg
   \y@m@veright@ifskew{#1}
    \vbox{
      \hrule height\y@linethick%
      \hbox{\ify@stdtext\y@abcbox#1\else\y@mathabcbox#1\fi\y@vr}
      \hrule height\y@linethick}\vskip-\y@linethick
  \else
   \y@m@veright@ifskew{#1}
    \vbox{
      \hrule height\y@linethick%
      \hbox{\y@abcboxes #1.\y@vr}%
      \hrule height\y@linethick}\vskip-\y@linethick
  \fi}
\def\young(#1){%
  \y@setdim%
  \hskip\y@interspace%
  \y@lastargtest#1,
  \ifmmode\ify@vcenter\vcenter\fi\fi{%
  \vbox{\offinterlineskip
    \ify@islastarg\y@nelineabcboxes{#1}%
    \else\y@ungabc(#1)%
    \fi}}\hskip\y@interspace}
\def\y@ungabc(#1,#2){%
  \y@nelineabcboxes{#1}%
  \y@lastargtest#2,
  \ify@islastarg\y@nelineabcboxes{#2}%
  \else\y@ungabc(#2)%
  \fi}
\catcode`\@12\relax
 
\psset{framearc=0.0,framesep=0.0truecm,nodesep=3pt,unit=0.75truecm,%
arrowsize=4pt 2,arrowinset=0.4}

\newcounter{mycnt}

\def\CO{\Delta} 
\def\co{\delta}

\def\antip{{\sf S}}

\def\ip{\ast}

\def\la{\langle}
\def\ra{\rangle}

\def\!{\kern -0.15ex}
\def\antip{\textsf{S}}

\def\CharGL{{\sf Char-GL }}

\def\piprod{\raisebox{0.2 ex}{${\scriptstyle \odot}$}\kern .2ex}
\def\ov{\overline}
\def\1{\ov{1}}
\def\2{\ov{2}}
\def\3{\ov{3}}

\begin{document}

\title[Algebraic Random Walks]
{Algebraic random walks in the setting of symmetric functions} 

%    author one information
%\thanks{}
\author{Peter D. Jarvis}
\address{P D Jarvis,  %
School of Mathematics and Physics, University of Tasmania, 
Private Bag 37, GPO, Hobart Tas 7001, Australia}
\email{Peter.Jarvis@utas.edu.au}
\author{Demosthenes Ellinas}
\address{D Ellinas,  %
Division of Mathematics, Technical University of Crete,
Polytechnioupolis GR-731 00 Chania Crete Greece}
\email{Ellinasd@gmail.com}

\subjclass[2000]{Primary 16W30; Secondary 05E05; }
%%16W30 Coalgebras, bialgebras, Hopf algebras [See also 16S40, 57T05];
%%      rings, modules, etc. on which these act
%%05E05 Symmetric functions
%%33D52 Basic orthogonal polynomials and functions associated with
%%      root systems (Macdonald polynomials, etc.)
%%11E57 Classical groups [See also 14Lxx, 20Gxx]
%%43A40 Character groups and dual objects

\keywords{plethysm, Hopf algebras, random walks,
algebraic combinatorics}

\date{July 2012}

\begin{abstract}
Using the standard formulation of algebraic random walks (ARWs) via coalgebras, we consider ARWs for co-and Hopf-algebraic structures in the ring of symmetric functions. These derive from different types of products by dualisation, giving the dual pairs of outer multiplication and outer coproduct, inner multiplication and inner coproduct, and symmetric function plethysm and plethystic coproduct. Adopting standard coordinates for a class of measures (and corresponding distribution functions) to guarantee positivity and correct normalisation, we show the effect of appropriate walker steps of the outer, inner and plethystic ARWs. If the coordinates are interpreted as heights or occupancies of walker(s) at different locations, these walks introduce translations, dilations (scalings) and inflations of the height coordinates, respectively.

\end{abstract}

\maketitle

%{\small\tableofcontents}
%%%%%%%%%%%%%%%%%%%%%%%%%%%%%%%%%%%%%%%%%%%%%%%%%%%%%%%%%%%%%%%%%%%%%%%%%%%
\section{Introduction and motivation}

The subject of algebraic random walks (ARWs) is a development of recent interest in the subject of `quantum probability' \cite{accardi:1981tqp,accardi:2000qphs}. The latter has origins in the foundations of quantum mechanics, but more recently has been influenced by trends towards noncommutative probability. ARWs 
use a very general algebraic framework \cite{meyer:1995quantum}, in formulating a variety of stochastic processes
such as Markov and L\'{e}vy processes \cite{barndorff:2006quantum}, via a Hopf algebraic formulation \cite{majid:2000foundations}.
Early studies used commutative function algebras and $q$-extensions to study Brownian motion random walks 
\cite{majid:1993qrwtr}; and more recently ARWs have been formulated for a variety of other
(co)algebraic structures such as the Weyl-Heisenberg algebra and $q$-deformations thereof \cite{ellinas:2001qdas}, as well as Lie algebraic enveloping algebras \cite{biane:1991qrwd}, and anyonic algebras. Interest in these studies is in the relationship of asymptotic limits to known (classical) stochastic processes, and also to some new types which have been shown to have behaviour which is significantly non-classical, such as variances linear in the step number rather than its square root, for example. For references and recent work see 
\cite{ellinas:2001qdas,ellinas:tsohantjis:2001bmsl,ellinas:tsohantjis:2003rwdsla,ellinas:2005aqrw}.
For a review of the related topic of quantum random walks we refer to Konno \cite{konno:2008quantum}.

The purpose of this work is to exploit the universal symmetric function ring $\Lambda(X)$ as an arena for ARWs.  
As is well known, $\Lambda(X)$ can be formulated as a Hopf algebra \cite{geissinger:1977a,zelevinsky:1981b} due to its intimate connection with character theory for the classical groups \cite{weyl:1930a}. In fact, several algebraic and co-algebraic structures coexist \cite{fauser:jarvis:2003a,fauser:jarvis:king:2012b} which means that there are rich possibilities for defining different types of ARWs. Specifically, we shall consider the outer Hopf algebra, the inner coalgebra, \cite{fauser:jarvis:2003a} and a plethysm coalgebra \cite{fauser:jarvis:king:2010a,fauser:jarvis:king:2012b}; these are all naturally defined by dualising standard products or pairings, namely the outer and inner symmetric function products, and the plethysm product, respectively (for notation see below); all can potentially lead to ARWs or composite variations on ARWs, and this will turn out to be the case. 

The main point of comparison for the symmetric function algebraic random walks which we shall describe, will be that of a random walk on a line. This is to be viewed in a discretized limit as in its classical ARW treatment by Majid \cite{majid:1993qrwtr,majid:1994random}. We will have discrete walker locations (in this case belonging to ${\mathbb N}$), and walk steps implemented by rearranging `height' coordinates representing the walker occupancy of the locations. We shall see that the different symmetric function coproducts involved in the convolutions operative in the time steps, lead to different classes of evolution. The main purpose of this paper is to point out that
these may be significantly richer than the standard case. Not only can local \emph{translations} (additive changes to heights), but also local \emph{scalings} (multiplicative changes to the heights) be generated; there are also, global \emph{inflations} of the sequence of height coordinates (for example, decimation, whereby heights are reallocated at intervals of 10 units, leaving intervening zeroes).

In the present note we do not consider these new possibilities in depth, but restrict ourselves simply to an identification or enumeration of the various cases of symmetric function ARWs (consistent with constraints such as positivity and normalisation). We defer investigations of asymptotic behaviour, continuum limits such as master or diffusion equations and the like, as well as interesting questions like identifying stationary states of some of the non-standard evolutions, to later detailed studies.
Concluding remarks on the interpretation of our findings, and some comparisons with other approaches into probability measures and Markov processes on symmetric functions, are given in \S \ref{sec:Conclusions}.

To complete this introductory section, we briefly review the technical framework of the ARW formalism
\cite{majid:1993qrwtr,majid:1994random,majid:2000foundations}.

%%%%%%%%%%%%%%%%%%%%%%%%%%%%%%%%%%%%%%%%%%%%%%%%%%%%%%%%%%%%%%%%%%%%%%%%%%%
\subsection{The ARW formalism.}
\label{subsec:ARWformalism}
We work in the framework of quantum probability spaces \cite{meyer:1995quantum}, although our subsequent application to standard symmetric functions will be in a commutative context. Generically, random variables will be appropriate linear functionals on some operator ${}^*$-algebra $H$, and probability measures are provided by states, that is, suitable normalised linear functionals.  We are concerned here to develop the consequences of additional co-algebraic structure carried by such $H$. 

Consider then a unital algebra $H$ with a co-unital, coassociative coalgebra $(H, \Delta)$, in which the multiplicative unit is group-like,
$\Delta(\eta)=\eta\otimes \eta$. We study 
linear functionals $\phi:H\rightarrow {\mathbb C}$ (or 1-cochains), and their convolution algebra.  Each such $\phi$ defines an algebra endomorphism $T_\phi: H \rightarrow H$, through
\begin{align}
T_\phi (f) = & \, \phi \otimes {\sf Id} \circ \Delta (f), \quad f \in H.
\nonumber
\end{align}
It follows from the counit axiom, $(\varepsilon \otimes {\sf id}) \circ \Delta = {\sf id} = ({\sf id}\otimes {\varepsilon}) \circ \Delta$, that 
$\varepsilon\big(T_\phi (f)\big) = \phi(f)$
so that $\phi$ can be freely recovered from its dual description $T_\phi$.
Finally $H$ becomes a probability space by giving a selected state, that is, a normalised, positive definite linear functional$\rho$ such that $\rho (\eta) = 1$ and $\rho(f^*f) \ge 0$. We can think of such $\rho$ as being associated with the integral over a density $\rho$,
$\rho(f) =  \int \rho \cdot f$, or, as below, via a scalar product with a density operator $\rho(f) = \langle \rho \mid f \rangle$.

The machinery of the stochastic process is now that we evolve random variables $f$ by the operation $T_\phi$ of convolution by $\phi$. Dually, we can regard the evolution as a random walk, taking place on the state itself by defining the pull-back $T_\phi^*$:
\begin{align}
%\label{}
T^*_\phi(\rho)\big(f \big):= & \, \rho\big(T_\phi(f)\big)  \nonumber
\end{align}
Evaluating the normalisation of the evolved density we require
\begin{align}
%\label{}
1= T^*_\phi(\rho)\big(\eta \big)  = & \,  \rho\big(T_\phi(\eta)\big) \nonumber\\
 = & \, \rho\big(\phi \otimes{\sf Id} \circ \eta \otimes \eta\big) \nonumber \\
 = &\,  \phi(\eta) \rho\big(\eta \big) \nonumber
\end{align}
where the group-like coproduct of $\eta$ has been used. Thus we require $\rho$ to be acted on by $T_\phi$ for normalised $\phi$, in order to retain the normalisation under the evolution. In general we have
\begin{align}
%\label{}
T^*_\phi(\rho)\big(f \big)= & \,\rho\big(T_\phi(f)\big)   \nonumber \\
= & \, \rho(f_{(2)}) \phi(f_{(1)}) \equiv \phi \star \rho (f). \nonumber 
\end{align}
Note similarly that
\begin{align}
T_\psi\big(T_\phi\big(f\big)\big) = & \, T_\psi\big(\phi(f_{(1)}) f_{(2)}\big) \nonumber \\
= & \, \phi(f_{(1)}) \psi(f_{(21)}) f_{(22)}; \nonumber \\
\mbox{compare} \qquad \qquad 
T_{\psi \star \phi}\big(f\big) = & \,\psi\star \phi(f_{(1)}) f_{(2)} \nonumber \\
= & \, \psi(f_{(11)}) \phi(f_{(12)}) f_{(2)}, \nonumber
\end{align}
and so, by coassociativity, we have $ T_\psi \circ T_\phi= T_{\psi \star \phi}$ -- that is, the endomorphisms with composition
form a semigroup which is homomorphic to the random variables (operators in $H$) under convolution. 

For positivity of the transformed state we note
\begin{align}
T^*_\phi(\rho)\big(f f^* \big)    =&\,  \rho\big(T_\phi(f f^*)\big) \nonumber\\
= & \,   \rho\big(\phi \otimes{\sf Id}(\Delta(f) \cdot \Delta (f^*))\big) \nonumber\\
= & \, \sum \phi(f_{(1)}f^*_{(1)'})  \rho\big(f_{(2)}f^*_{(2)'}\big)\nonumber
\end{align}
where the $'$ notation signifies an independent summation over the coproduct parts of $f$ and $f^*$. 

In the symmetric function context (see \S \ref{sec:SymmFns} below), we shall be often be working over ${\mathbb R}$, and positivity involves $f^2$ rather than $f f^*$ (the conjugation defined by ${}^*$ here is not to be confused with the ${}^*$ used to designate duals above). Moreover, 
the state will be given by taking a scalar product with respect to a specific element, $\rho(f) = \langle \rho\mid f\rangle_H$.
In this case, a sufficient condition for positivity is provided by one additional, quite natural assumption. Since this will be used repeatedly in the sequel, we state it here as a formal lemma:
\begin{quotation}
For a scalar product $\langle \cdot \mid\cdot \rangle_H : H \otimes H \rightarrow {\mathbb C}$ which is compatible with the multiplication in $H$ (see below), then the linear functional $\rho: H \rightarrow {\mathbb C}$ defined by $\rho(\cdot) = \langle \rho\mid \cdot \rangle_H$
is positive, $\langle \rho \mid f^2 \rangle \ge 0$ for arbitrary $f\in H$ with  $\rho(f) \in {\mathbb R}$, provided $\rho$ is group-like, $\Delta(\rho) = \rho\otimes \rho$, or is a positive (convex) combination of group like elements.
\end{quotation}
The property is trivially verified using the compatibility property, 
\[
\langle \rho \mid f^2 \rangle_H = \langle \Delta(\rho)  \mid f\otimes f \rangle_H 
= \langle \rho\otimes \rho  \mid f\otimes f \rangle_H  \equiv \langle \rho \mid f \rangle_H^2 \ge 0.
\]
Positivity (and normalisation) are clearly maintained by the more general possibility of a positive (convex) combination of group-like elements.

The property of scalar product-product compatibility, $\langle x \mid yz \rangle = \langle \Delta(x) \mid y \otimes z \rangle$, is formally called a Laplace pairing \cite{rota:stein:1994b,rota:stein:1994a}. Where the coproduct is defined by the right-hand side, this is of course built-in. However, 
in situations such as the symmetric function setting, where two or more products and coproducts coexist, the Laplace pairing property can take several different forms.  Note that the discussion does not require the bialgebra axiom of product-coproduct compatibility, $\Delta(xy) = \Delta(x)\Delta(y)$, but where available this can of course be used to advantage.  Again, if the coproduct is defined by induction starting from primitive elements, then this is built-in; however in the general setting, only certain of the different forms of the bialgebra property may hold. For the purposes of the present paper, further assumptions about bialgebras are not required.

Finally, algebraic random walks are constructed via repeated evolutionary steps from a starting state $\rho$, (the underlying density  itself), under $T_\phi$, or equivalently, by evaluating successive convolution powers $\phi \star \rho$, $\phi \star \phi \star \rho$, $\cdots$, $(\star^n \phi) \star \rho$. In looking at relations with stochastic processes, for example, exponential forms such as $
\exp_\star(\phi) = \lim_{n\rightarrow \infty} \star^n(1 + \phi/n) $, can be examined to deduce generalised random walks, and to study continuum limits and (non) classical behaviours \cite{majid:1993qrwtr,majid:1994random,ellinas:2001qdas,ellinas:tsohantjis:2001bmsl,ellinas:tsohantjis:2003rwdsla,ellinas:2005aqrw}. \\[.2cm]

In this paper we apply the above machinery in the context of a well known structure, the Hopf algebra of symmetric functions, with the aim of elucidating the possibilities of ARWs in this setting. 

The structure of this paper is as follows. The co-algebraic set-up for defining ARWs has been briefly reviewed in subsection \S \ref{subsec:ARWformalism} above. \S \ref{sec:SymmFns} below provides a technical summary of the relevant algebraic properties of the ring of symmetric functions, which are needed to implement these definitions. \S \ref{sec:SymmFnARW} treats the three types of ARWs derived from the three (dualised) coproducts identified in the structural data of the symmetric functions, namely the outer-ARW (\S \ref{subsec:OuterARW}), the inner ARW (\S \ref{subsec:InnerARW}), and the plethystic-ARW (\S \ref{subsec:PlethysticARW}). In each case, a selected class of probability measures (states) is identified, under which the ARW transformations appear natural and can be given explicitly for a single step (see for example the related work of \cite{brouder:fauser:frabetti:oeckl:2002a} for closed-form 
expressions for $n$-fold convolutions). For each of our ARW cases, a summary statement in the form of a technical lemma is provided. 
Concluding remarks on the interpretation of our findings, and some comparisons with other approaches into probability measures and Markov processes on symmetric functions, are given in \S \ref{sec:Conclusions}.

%%%%%%%%%%%%%%%%%%%%%%%%%%%%%%%%%%%%%%%%%%%%%%%%%%%%%%%%%%%%%%%%%%%%%%%%%%%
\section*{Acknowledgement}
PDJ acknowledges the hospitality of the Department of Mathematics, Section of Sciences, and colleagues at the Technical University of Crete during a research visit. PDJ acknowledges correspondence and collaboration with Bertfried Fauser and Ronald King on work related to this paper. The authors thank Bertfried Fauser for useful suggestions on the manuscript and for pointing out several references. Part of this work was completed under an Australian senior Fulbright scholarship (Department of Statistics, University of California Berkeley, and Department of Physics, University of Texas at Austin) and hosts and colleagues at these institutions are thanked for their support.
%%%%%%%%%%%%%%%%%%%%%%%%%%%%%%%%%%%%%%%%%%%%%%%%%%%%%%%%%%%%%%%%%%%%%%%%%%%
%\newpage
\section{The Hopf algebra of symmetric functions}
\mbox{}

We introduce and briefly review the salient features of symmetric functions needed for our purposes.

\label{sec:SymmFns}
\subsection{The ring of symmetric functions $\Lambda(X)$.}
We consider characters of finite-dimensional polynomial (tensor) representations of the complex group $GL(n)$ of $n\times n$ nonsingular matrices, extended to a ring over ${\mathbb Z}$ including formal subtraction as well as addition and multiplication of characters. In the inductive limit this object \CharGL  is isomorphic to the ring of symmetric functions $\Lambda(X)$ on an alphabet $X$ of countably many variables  $\{ x_1,x_2,x_3,\cdots \}$. $\Lambda(X)$ has a canonical basis involving irreducible $GL(n)$ characters, the Schur or $S$-functions $\{ s_\lambda \}_\lambda$ where $\lambda$ is an integer partition. Below we turn to the attributes of $\Lambda(X)$ which play a  
crucial role in our development, namely its algebraic and Hopf-algebraic structures \cite{fauser:jarvis:2003a}. Where no ambiguity arises, $\Lambda(X)$ will occasionally be referred to simply as $\Lambda$ in what follows. For our applications it is convenient to work over ${\mathbb R}$. Firstly we give some notational preliminaries.
\subsection{Partitions and Young diagrams}

Our notation follows in large part that of~\cite{macdonald:1979a}.  
Partitions are specified by lower case Greek letters. 
If $\lambda$ is a partition of $n$ we write $\lambda\vdash n$, and 
$\lambda=(\lambda_1,\lambda_2,\ldots,\lambda_n)$ is a sequence of 
non-negative integers $\lambda_i$ for $i=1,2,\ldots,n$
such that $\lambda_1\geq\lambda_2\geq\cdots\geq\lambda_n\geq0$, 
with $\lambda_1+\lambda_2+\cdots+\lambda_n=n$. The partition
$\lambda$ is said to be of weight $|\lambda|=n$ and length
$\ell(\lambda)$ where $\lambda_i>0$ for all $i\leq\ell(\lambda)$ 
and $\lambda_i=0$ for all $i>\ell(\lambda)$. In specifying $\lambda$
the trailing zeros, that is those parts $\lambda_i=0$, are often 
omitted, while repeated parts are sometimes written in exponent form
$\lambda=(\cdots,2^{m_2},1^{m_1})$ where $\lambda$ contains $m_i$
parts equal to $i$ for $i=1,2,\ldots$. For each such partition,
$n(\lambda)=\sum_{i=1}^n (i-1)\lambda_i$ and 
$z_\lambda=\prod_{i\geq1} i^{m_i}\, m_i!$.
\medskip

Each partition $\lambda$ of weight $|\lambda|$ and length $\ell(\lambda)$
defines a Young or Ferrers diagram, $F^\lambda$, consisting of $|\lambda|$ 
boxes or nodes arranged in $\ell(\lambda)$ left-adjusted rows of lengths 
from top to bottom $\lambda_1,\lambda_2,\ldots,\lambda_{\ell(\lambda)}$ 
(in the English convention). The partition $\lambda'$, conjugate to 
$\lambda$, is the partition specifying the column lengths of $F^\lambda$ read 
from left to right. The content of the box $(i,j)\in F^\lambda$, that is in 
the $i$th row and $j$th column of $F^\lambda$, is said to have content 
$c(i,j)=j-i$ and hook length $h(i,j)=\lambda_i+\lambda'_j-i-j+1$. 

By way of illustration, if $\lambda=(4,2,2,1,0,0,0,0,0)=(4,2,2,1)=(4,2^2,1)$ then
$|\lambda|=9$, $\ell(\lambda)=4$, $\lambda'=(4,3,1^2)$,
\begin{equation}
  F^\lambda=F^{(4,2^2,1)} =\ \raisebox{-0.8cm}{\yng(4,2,2,1)}\ \ \hbox{and}\ \
  F^{\lambda'}=F^{(4,3,1^2)} =\ \raisebox{-0.8cm}{\yng(4,3,1,1)} \ \ .
\end{equation}
The content and hook lengths of $F^\lambda$ are specified by
\begin{equation}
  \raisebox{-0.8cm}{\young(0123,\10,\2\1,\3)}\ \qquad \hbox{and}\qquad \ 
  \raisebox{-0.8cm}{\young(7521,42,31,1)}\ \ ,
\end{equation}
where $\ov{m}=-m$ for all $m$. 
In addition, $n(4,2^2,1)=0\cdot4+1\cdot2+2\cdot2+3\cdot1=9$ and 
$z_{(4,2^2,1)}=4\cdot2^2\cdot1\cdot1!\cdot2!\cdot1!=32$.

\subsection{The ring $\Lambda(X)$ and Schur functions}

There exist various bases of $\Lambda(X)$ as described 
in~\cite{macdonald:1979a}: the monomial symmetric functions 
$\{m_\lambda\}_\lambda$, the complete
symmetric functions $\{h_\lambda\}_\lambda$, the elementary symmetric
functions $\{e_\lambda\}_\lambda$, the power sum symmetric functions
$\{p_\lambda\}_\lambda$ and the Schur symmetric 
functions $\{s_\lambda\}_\lambda$. Three of these bases are multiplicative,
with $h_{\lambda}=h_{\lambda_1}h_{\lambda_2}\cdots h_{\lambda_n}$, $e_{\lambda}=e_{\lambda_1}e_{\lambda_2}\cdots e_{\lambda_n}$ and $p_{\lambda}=p_{\lambda_1}p_{\lambda_2}\cdots p_{\lambda_n}$.
Of the relationships between the various bases, we just mention at this stage the transitions corresponding to the Frobenius 
homomorphism
\[
    p_\mu(X) =\sum_{\lambda\vdash n}\ \chi^\lambda_\mu\ s_\lambda(X)
    \quad\hbox{and}\quad
   s_\lambda(X) = \sum_{\mu\vdash n}\ z_\mu^{-1}\ \chi^\lambda_\mu\ p_\mu(X)\,,
\label{Eq-p-s}
\]
where $|\mu| =n=|\lambda|$ and $\chi^\lambda_\mu$ is the character of the irreducible representation
of the symmetric group ${\mathfrak S}_n$ specified by $\lambda$ in the conjugacy class
specified by $\mu$. These characters satisfy the orthogonality conditions
\[
    \sum_{\mu\vdash n}\ z_\mu^{-1}\ \chi^\kappa_\mu\ \chi^\lambda_\mu\ = \ \delta_{\kappa,\lambda}
    \quad\hbox{and}\quad
   \sum_{\lambda\vdash n}\ z_\mu^{-1}\ \chi^\lambda_\alpha\ \chi^\lambda_\beta\ = \ \delta_{\alpha,\beta}\,.
\label{Eq-chi-orth}
\]

The significance of the Schur function basis lies in the fact that with 
respect to the usual Schur-Hall scalar product $\langle \cdot \,|\, \cdot \rangle_{\Lambda(X)}$
on $\Lambda(X)$ we have
\[
       \langle s_\mu(X) \,|\, s_\nu(X) \rangle_{\Lambda(X)} = \delta_{\mu,\nu}\,. 
\label{Eq-scalar-prod-s}
\]
From above it follows that
\[
       \langle p_\alpha(X) \,|\, p_\beta(X) \rangle_{\Lambda(X)} = z_\alpha \delta_{\alpha,\beta}\,. 
\label{Eq-scalar-prod-p}
\]

In what follows we shall make considerable use of several infinite series,
which serve as generating functions for some of 
the bases of the ring of symmetric functions mentioned above. The most important of these are the mutually inverse
pair defined by
\begin{align}
   M(t;X) &= \prod_{i\geq1} (1-t\,x_i)^{-1} =  \sum_{m\geq0}\ h_m(X)\,t^m = \exp\big( \sum_{m>0} p_m(X)\frac{t^m}{m}\big)\,;  \nonumber \\ %\label{Eq-M}
   L(t;X) &= \prod_{i\geq1} (1-t\,x_i) =  \sum_{k\geq0}\ (-1)^m\, e_m(X)\,t^m\,,  \nonumber %\label{Eq-L}
\end{align} 
where, as Schur functions\footnote{In Macdonald's notation and $\lambda$-ring 
notation $M(t;X)=H(t)=\sigma_t(X)$ and $L(t;X)=E(-t)=\lambda_{-t}(X)$.}, $h_m(X)=s_{(m)}(X)$ and $e_m(X)=s_{(1^m)}(X)$ ; the power sum generators are by definition $p_m = \sum_i x_i^m$.
In the case $t=1$ we write simply $M(1;X)=M(X)$ and $L(1;X)=L(X)$.

\subsection{Algebraic properties of $\Lambda(X)$}

The ring, $\Lambda(X)$, of symmetric functions over $X$ has a Hopf algebra structure
%The ring, $\Lambda(X)$, $\bigwedge(X)$, of symmetric functions over $X$ has a Hopf algebra structure
(\cite{geissinger:1977a,zelevinsky:1981b}; see also \cite{thibon:1991a,thibon:1991b,scharf:thibon:1994a} ), with 
two further algebraic and two coalgebraic operations. For notation and
basic properties we refer for example 
to~\cite{fauser:jarvis:2003a,fauser:jarvis:king:wybourne:2005a}, and references therein (for related aspects see also
\cite{fauser:jarvis:king:2010a,fauser:jarvis:king:2012a,fauser:jarvis:king:2012c,fauser:jarvis:king:2012b}). 
In the interest of typographical simplicity, the symbol $X$ for the 
underlying alphabet is often suppressed unless specifically required. 

We indicate outer products on $\Lambda$ either by $m$, or with infix notation 
using juxtaposition. Inner products are denoted either by $\textsf{m}$ or as infix 
by $\ip$, while plethysms (compositions) are denoted by $\circ$ or 
by means of square brackets $A[B]$.
With respect to the Schur-Hall scalar product, the outer product Hopf algebra
  $\Lambda$ is isomorphic to its dual $\Lambda^*$.
In the same way, the inner product dualizes to a coproduct, and in \cite{fauser:jarvis:2003a} it is established that the only 
bialgebras are between the outer product and inner coproduct, and inner product and outer coproduct, respectively.  By contrast, the dualization of the more problematic plethystic product
simply yields a coproduct \cite{fauser:jarvis:king:2012b}.

Explicitly, the coproduct maps corresponding to the three bilinear products are specified by $\Delta$ for the outer coproduct, $\delta$ 
for the inner coproduct, and $\nabla$ for the plethysm coproduct\footnote{We do not treat inner plethysm \cite{scharf:thibon:1994a} in the present work.}. 
In Sweedler
notation the action of these coproducts is distinguished by means of different brackets, 
round, square and angular, around the Sweedler indices.  The coproduct coefficients themselves are 
obtained from the products by duality using the Schur-Hall scalar 
product and the self-duality of $\Lambda(X)$. For all $A,B\in\Lambda(X)$
with $\Lambda^+(X) = \ker \varepsilon$ (see below), we have then
\begin{align}
m(A\otimes B) &= AB 
&&&
\CO(A) &= A_{(1)}\otimes A_{(2)} &&&  \textrm{Hopf algebra,} \nonumber \\
{\sf m}(A\otimes B) &= A\ip B 
&&&
\co(A) &= A_{{[}1{]}}\otimes A_{{[}2{]}}&&&  \textrm{(bialgebra),} \nonumber \\
A\circ B &= A[B] 
&&&
\nabla(A) &= A_{<1>}\otimes A_{<2>} &&& \textrm{coalgebra on $\Lambda^+$.} \nonumber  
\end{align}

In terms of the Schur function basis $\{s_\lambda\}_{\lambda\vdash
n,n\in\mathbb{N}}$ the product and coproduct maps give
rise to the particular sets of coefficents specified as follows:
\begin{align}
s_\mu s_\nu &= \sum_\lambda c^\lambda_{\mu,\nu}s_\lambda 
&&&
\CO(s_\lambda) 
  &= s_{\lambda_{(1)}}\otimes s_{\lambda_{(2)}}
   =\sum_{\mu,\nu}c^\lambda_{\mu,\nu} s_\mu\otimes s_\nu
   \nonumber \\
s_\mu \ip s_\nu &= \sum_\lambda g^\lambda_{\mu,\nu}s_\lambda 
&&&
\co(s_\lambda) 
  &= s_{\lambda_{[1]}}\otimes s_{\lambda_{[2]}}
   =\sum_{\mu,\nu}g^\lambda_{\mu,\nu} s_\mu\otimes s_\nu
   \nonumber \\
s_\mu[s_\nu] &= \sum_\lambda p^\lambda_{\mu,\nu}s_\lambda 
&&&
\nabla(s_\lambda) 
  &= s_{\lambda_{\la 1\ra}}\otimes s_{\lambda_{\la 2\ra}}
   =\sum_{\mu,\nu}p^\lambda_{\mu,\nu} s_\mu\otimes s_\nu
   \nonumber
\end{align}
The significance of the various products is as follows.
The outer product, simply pointwise multiplication of symmetric functions, 
arises at the group character level from the reduction of a tensor product of
representations of the general linear group. The $c^\lambda_{\mu,\nu}$ are famous Littlewood-Richardson coefficients.
By contrast the structure coefficients for the inner product $g^\lambda_{\mu,\nu}$ are the Kronecker 
coefficients corresponding to the reduction
of tensor products of representations of the symmetric group, 
lifted to Schur functions via the above Frobenius mapping.
Finally, plethysm is the word used by Littlewood to describe functional composition of symmetric functions.
The plethysm coefficients $p^\lambda_{\mu,\nu}$ arising from the symmetric function 
$s_\mu[s_\nu]$, correspond at the character level to taking a $\mu$-symmetrised tensor power,
associated under the symmetric group with some partition $\mu$, of a representation
of the general linear group (in this case with irreducible character $s_{\nu}$),
and expressing the result in terms of irreducibles (note that we require $|\lambda| \ge 1$).

All these coefficients 
are non-negative integers. The Littlewood-Richardson coefficients
can be obtained for example, by means of the Littlewood-Richardson 
rule~\cite{littlewood:1940a} or the hive model~\cite{buch:2000a}. The inner product Kronecker 
coefficients may determined directly from characters of the symmetric group 
or by exploiting the Jacobi-Trudi identity and the 
Littlewood-Richardson rule~\cite{robinson:1961a}, while plethysm coefficients
have been the subject of a variety methods of calculation~\cite{littlewood:1940a,
robinson:1961a,chen:garsia:remmel:1984a} and symbolic evaluation \cite{schur,schurfkt:2003a}.
Note that the above sums are finite, since
\begin{align}
c^\lambda_{\mu,\nu} &\ge0 
  \qquad\textrm{iff}\quad \vert\lambda\vert=\vert\mu\vert +\vert\nu\vert\,;
  \nonumber \\
g^\lambda_{\mu,\nu} &\ge0 
  \qquad\textrm{iff}\quad \vert\lambda\vert=\vert\mu\vert =\vert\nu\vert\,;
  \nonumber \\
p^\lambda_{\mu,\nu} &\ge0 
  \qquad\textrm{iff}\quad \vert\lambda\vert=\vert\mu\vert\,\vert\nu\vert\,.
  \nonumber
\end{align}

The Schur-Hall scalar product may be used to define skew products $s_{\lambda/\mu}$
of Schur functions through the identities
\[
     c^\lambda_{\mu,\nu}= \la\, s_\mu\, s_\nu \,|\, s_\lambda\,\ra = \la\, s_\nu\,|\, s^\perp_\mu (s_\lambda)\, \ra
          = \la \, s_\nu\,|\, s_{\lambda/\mu}\, \ra\,,
\label{Eq-def-skew}
\]
so that 
\[
         s_{\lambda/\mu} =  \sum_\nu\ c^\lambda_{\mu,\nu}\ s_\nu\,.
\label{Eq-skew}
\]    

For the outer product and coproduct bialgebra we have a unit $\eta$, 
a counit $\varepsilon$ and (completing the Hopf structure) an antipode $\antip$ such that
\[
     \eta(1)=s_{(0)}\,;\qquad \varepsilon(s_\lambda)=\delta_{\lambda,(0)}\,; \qquad \antip(s_\lambda) = (-1)^{|\lambda|} s_{\lambda'}\,,
\label{Eq-antip}
\]  
where we have extended the notation to include the unit symmetric function for the trivial (empty) partition $(0)$, which plays the role of the multiplicative unit $1_m$. For the inner product algebra and coalgebra we have a unit $\eta_{\sf m}=1_{\sf m}$, and
co-unit $\varepsilon^{\sf m}$ given by\footnote{The inner algebra and coalgebra do not form a bialgebra, and there is no antipode.}
\[
1_{\sf m} \equiv M , \qquad \varepsilon^{\sf m} \equiv \langle M \mid \cdot \, \rangle .
\]
%%%%%%%%%%%%%%%%%%%%%%%%%%%%%%%%%%%%%%%%%%
%%%%%%%%%%%%%%%%%%%%%%%%%%%%%%%%%%%%%%%%%%
%%%%%%%%%%%%%%%%%%%%%%%%%%%%%%%%%%%%%%%%%%
%%%%%%%%%%%%%%%%%%%%%%%%%%%%%%%%%%%%%%%%%%
\section{ARWs and symmetric functions}
\mbox{}
\label{sec:SymmFnARW}
We wish to apply the ARW construction to the ring of symmetric functions itself, using its Hopf and co-algebraic structures, and to explore the nature of the resulting random walks. While it can be expected that the results will be analogous to those for the standard random walks on the line, the nature of the Hopf- and co-algebras involved is much richer, and provides fertile ground for new types of walks, diffusions and master equations.

In this spirit of `experimental' investigation, we take in turn the successive cases of the outer Hopf algebra $(\Lambda, \Delta)$, the inner coalgebra $(\Lambda, \delta)$, and finally the plethysm coalgebra $(\Lambda, \nabla)$, denoting the corresponding ARWs as outer ARWs, inner ARWs and plethystic ARWs, respectively.
As mentioned in the introduction, we can guarantee positivity by taking the state, $\rho$, to be a group-like element of an appropriate coproduct. This will be explored on a case-by-case basis, and for some discussions more general states are considered.

%%%%%%%%%%%%%%%%%%%%%%%%%%%%%%%%%%%%%%%%%%
\subsection{Outer ARW for $(\Lambda, \Delta)$}
\label{subsec:OuterARW} Group-like elements of the outer symmetric function Hopf algebra have to be constructed via formal series, or technically ring extensions $\Lambda(X)[[t]]$, of which the series $M(X,t)$ and $L(X,t)$ of the previous section are examples. One way to parametrise a general group-like element is to note that the power sums $p_n(X)$ are primitive elements with respect to the outer coproduct, namely
\[
\Delta p_n = p_n \otimes 1 + 1 \otimes p_n ,
\]
from which it follows immediately by the coalgebra property and product laws\footnote{It is worth noting that in a non (co)commutative setting, for example  $\Delta p_n = p_n \otimes q + q^{-1} \otimes p_n$, group-like elements can still be recovered by using an appropriate $q$-exponential, $\Delta\big(\exp_q(p_n)\big) = \exp_q(p_n) \otimes \exp_q(p_n)$.}
that $\exp(p_n)$ is group-like, $\Delta\big(\exp(p_n)\big) = \exp(p_n) \otimes \exp(p_n)$. In general we take a group-like series,
\begin{align}
\label{eq:Mdefn}
M(X;\{c_n\}) = & \, \exp\big(\sum_{n=1}^\infty p_n(X)c_n/n\big),
\end{align}
or slightly more flexibly using a product alphabet $XU = \{x_i u_j\}_{i,j \in {\mathbb N}}$,
\[
M(XU) = \prod_{i,j}\frac{1}{(1-x_iu_j)} = \exp\big(\sum_{n=1}^\infty p_n(X)p_n(U)/n\big) = \sum_\alpha s_\alpha(X)s_\alpha(U).
\]
The last form of $M$ is the so-called Cauchy identity, and the second follows directly from the series expansion of $\ln(1-x)$ as noted already in the previous definitions. The connection between these forms is that the $p_n(U)$ are regarded as a set of algebraically independent generators (of $\Lambda(U)$), and so can in principle be specialized to numerical values such that $p_n(U)\equiv c_n$. In practice, any symmetric functions in $U$ arising from manipulations involving $M(XU)$ must be expressed (using the Frobenius map) as polynomials in $p_n(U)$, from which in turn an expression polynomial in $c_n$ can be recovered (see below).  When the parameters $\{c_n\}_{n\in{\mathbb N}}$ are to be emphasized, we use the notation $M_c(X)$, whereas in $S$-function manipulations, we use $M(XU)$, with the understanding that the above specialization of power sums is to be implemented at the end of any calculations. 

We now assume that the action of the state $M$ on random variables (symmetric functions) proceeds via the Schur-Hall scalar product,
$M(f) := \langle M \mid f \rangle_\Lambda$, from which we check
\begin{align}
M(\eta)= & \, \langle M \mid \eta \rangle_\Lambda = 1, \qquad \nonumber \\
M(f^2) = & \,  M_{(1)}(f)M_{(2)}(f) = \langle M \mid f \rangle_\Lambda^2,
\nonumber
\end{align}
where $\Delta(M) = M \otimes M$ and the Laplace property of $\langle \cdot \mid \cdot \rangle_\Lambda$ with respect to the outer symmetric function product has been used \cite{fauser:jarvis:2003a}. If there is no ambiguity we refer to the scalar product simply as $\langle \cdot \mid \cdot \rangle$.  
For a general evaluation of our state $M$ on an arbitrary symmetric function we have, using the $S$-function basis for $f$, say $f(X) = \sum_\alpha
 f_\alpha s_\alpha(X)$, and the Cauchy identity for $M$,
\[
\label{eq:MfUform}
\langle M(XU) \mid f(X) \rangle= \langle \sum_\alpha s_\alpha(X)s_\alpha(U) \mid \sum_\beta f_\beta s_\beta(X) \rangle_{\Lambda(X)}
\equiv \sum_\alpha f_\alpha s_\alpha(U) \equiv f(U)
\]
where as mentioned above, the $s_\rho(U)$ have to be expressed in terms of the $c_n$. Concretely this is done by re-writing 
each $s_\alpha(X)$ as $S_\alpha(p)$, the polynomial in power sums defined by the right-hand side of the Frobenius mapping between Schur functions and power sum functions: that is, we define
$S_\alpha(c) := \sum_\beta z_\beta^{-1}\chi^\alpha_\beta c_\beta$ and the $c_\alpha$ are monomials in the $c_n$ built up multiplicatively, in the same way as the power sum functions,
$c_\alpha = c_{\alpha_1}c_{\alpha_2}\cdots$, consistent with the identification of $p_n$ with $c_n$.
%$S_\alpha(p) := \sum_\beta z_\beta^{-1}\chi^\alpha_\beta p_\beta$.
Thus the result of the evaluation is finally
\[
M_c(f) = \sum_\alpha {f}_\alpha S_\alpha(c) . %\equiv \sum_\alpha {f}_\alpha c_\alpha
\]

For the random walk evolution, we make the natural assumption that the linear forms $\phi$, and correspondingly the endomorphisms denoted in this section by $T{}^{out}_\phi$, just consist in taking the scalar product with specific symmetric functions $\phi$. We consider choices for $\phi$ momentarily:
\begin{align}
T{}^{out}_\phi(f) = & \, \langle \phi \mid f_{(1)} \rangle f_{(2)}; \nonumber \\
{T{}^{out}_\phi}^*\big(M_c\big) (f) = & \, \langle M(XU) \mid f_{(2)}\rangle \langle \phi \mid f_{(1)} \rangle \equiv \langle M(XU)\!\cdot\! \phi(X) \mid f\rangle \nonumber
\end{align}
where again the (left) Laplace property has been used\footnote{The last expression can be rearranged using commutativity and the property of the Cauchy series in the form, 
$\phi(X)\!\cdot\! M(XU) = \phi(U)^\perp M(XU)$. Either of these forms can be used as ${T{}^{out}_\phi}^*\big(M(XU)\big)$.}. 
Checking positivity by evaluation on a square, we have
\begin{align}
\langle M(XU)\phi(X) \mid f^2\rangle_{\Lambda(X)} = 
& \, \langle (M\phi)_{(1)} \mid f\rangle \langle (M\phi)_{(2)} \mid f\rangle \nonumber \\
= & \, \langle M\phi_{(1)} \mid f\rangle \langle M \phi_{(2)} \mid f \rangle \nonumber \\
\equiv & \, (\phi_{(1)}^\perp f)(U)\cdot (\phi_{(2)}^\perp f)(U).\nonumber
\end{align}
using, respectively, the Laplace property of the scalar product, the outer bialgebra product-coproduct axiom, and once again the group-like nature of the $M$ series. The last form derives from the definition of the skew as the adjoint of multiplication, and using again the reproducing property $\langle M(XU)\mid g(X) \rangle_{\Lambda(X)} = g(U)$ for any $g$ as noted above. 
Once again, a sufficient condition for positivity is in turn that $\phi$ should be group-like (or a convex combination, as below), say 
\[
\phi(X) = \exp\big(\left.\sum\right._{n=1}^\infty p_n(X) \varphi_n \big).
\]
Finally checking normalisation,
\begin{align}
{T{}^{out}_\phi}^*\big(M_c\big) (\eta)= & \, \langle M(XU) \mid T{}^{out}_\phi(\eta)\big\rangle_{\Lambda(X)} \nonumber \\
= & \, \langle \phi \mid \eta\big \rangle \langle M(XU) \mid \eta \rangle \equiv \phi(\eta)M_c(\eta) \nonumber 
\end{align}
(as already noted in the abstract setting). Thus ${T{}^{out}_\phi}^{*}\big(M_c\big)$ remains normalised provided $\phi$ itself is normalised.

A slightly more general form of $\phi$ (which induces a more general form of the state $M_c$ under evolution) is to consider a convex sum of such group-like elements:
\begin{align}
\label{eq:GeneralPhiForm}
\phi(X) = & \, \sum_{i=1}^k \lambda_i \exp\big(\left.\sum\right._{n=1}^\infty p_n(X) \varphi^i_n \big),
\qquad \sum_{i=1}^k \lambda_i =1, \qquad \lambda_i \ge 0.
\end{align}
(the convexity is required by normalisation; $k=1$ is the previous case). In summary, then, the state function $M_c \cong M(XU)$, coordinatised by the sequence $\{c_n\}_{n\in{\mathbb N}}$, is acted on by the flows generated by $\phi(X)$, also of exponential form, but parametrised instead by the sequences $\{ \varphi^i_n \}_{n\in{\mathbb N}}, \lambda_i, i=1,2,\cdots,k, \lambda_i \ge 0, \sum_i \lambda_i=1$. Obviously, the action of $\phi$ resulting from pointwise multiplication of exponentials, $\phi(X)M(XU)$,  simply adds the parameters in the exponential.  
This situation is similar to that of random walks on the line, but now the evolution is expressed in terms of symmetric function operations, as summarised by the following lemma:

\begin{quotation}
\emph{Outer product algebraic random walk on symmetric functions.}\\
Consider the parametrised state $M(X; \{c_n\}_{n\in {\mathbb N}})$ as in (\ref{eq:Mdefn}) above. Denote the sequence  $\{c_n\}_{n\in {\mathbb N}}$ by $c$, and the state by $M_c$. Under the outer-ARW evolution $T_\phi^{out}{}^*$ generated by the element $\phi(X)$ defined above, we have
%and similarly for $\{c_n+ \varphi^i_n\}_{n\in {\mathbb N}}$ $\cong$ $c+\varphi^i$, we have
\begin{align}
M_c \rightarrow & \, \sum_i \lambda_i M_{c+\varphi^i}. \nonumber
\end{align}
with the following tentative interpretation. The coordinates $\{c_n\}$ are the initial heights of a walker starting distribution. After each step there are translations, with probability $\lambda_i$, by the amounts $\{ \varphi^i_n\}$ in each coordinate. These translations accumulate multinomially with successive iterations of the convolution with $\phi$.
\end{quotation}

Finally we demonstrate the effect of the transformed state on an arbitrary random variable (symmetric function) $f= \sum_\alpha f_\alpha s_\alpha $. In (\ref{eq:MfUform}) it was shown that $M(f) = f(U)$. In order to encapsulate the specialization $p_n(U)=c_n$, we implement the transition between symmetric functions and power sums via the polynomials $S_\alpha(p)$
defined above. Thus the initial and transformed measures of $f$ are
simply the derived combinations of these polynomials and shifted forms, as follows:
\begin{quotation}
\emph{Measurement of symmetric functions}\\
The weight of a random variable (symmetric function) $f$ under the starting measure $M_c$ is simply $M_c(f) = \sum_\alpha f_\alpha S_\alpha(c)$. The transformed measure after evolution by $T^{out}_\phi$ involves the corresponding  convex sum of shifted $S_\alpha$ polynomials.
\[
T_\phi{}^{out}{}^*(M_c)\big(f\big) = \sum_i \lambda_i M_{c+\varphi^i}(f) = \sum_\alpha f_\alpha\big(\sum_i \lambda_i S_\alpha(c+\varphi^i)\big).
\]
\end{quotation}
 
A final aspect of the outer-ARW is to examine the effect of the active transformation on the random variables themselves, $f \rightarrow T^{out}_\phi(f)$. Fix $f$ to be a specific symmetric function, say a Schur function $f = s_\lambda(X)$. Then given the outer coproduct $s_\lambda(Y,X) = \sum_\alpha s_\alpha (Y) s_{\lambda/\alpha}(X)$,
and the reproducing property $\langle M(Y,U)\mid s_\alpha(Y)\rangle_{\Lambda(Y)} = s_\alpha(U)$ noted already, we establish the following:

\begin{quotation}
\emph{Outer-ARW action on symmetric functions}\\
The evolution of a given Schur function $s_\lambda(X)$ under the ARW generated by successive actions by elements $\phi(X), \psi(X), \cdots $ as in
(\ref{eq:GeneralPhiForm}) above is as follows:
\begin{align}
\label{eq:GLbranch}
s_\lambda(X) \rightarrow & \, \sum_i \lambda^i \big( \sum_\alpha s_{\lambda/\alpha}(X) S_\alpha(\varphi^i) \big)
\nonumber \\
\rightarrow & \, 
\sum_{i,j} \lambda^i \mu^j \big( \sum_{\alpha,\beta} s_{\lambda/\alpha \beta}(X) S_\alpha(\varphi^i)
S_\beta(\psi^j) \big) 
\rightarrow   \cdots 
\end{align}
with probabilities $\lambda^i, \mu^j, \cdots$ for the corresponding forms.
\end{quotation}

One of the main features of the foregoing is the group-like nature of the state $\rho$. The choice
$M(X; \{c_n\}_{n\in {\mathbb N}})$ can be modified, for example to series of the form
\[
\prod_i \frac{1}{(1- f(x_i)u_j)} 
\]
for some polynomial $f(x)$, without losing the group-like coproduct. While the action on random variables (symmetric functions) will be quite different from the standard case under the corresponding $T^o_\phi $ mappings, their convolution (generated by the series outer product) will have similar features to the minimal case already discussed, and will not be further considered.  
It is also possible to relax the condition of having a group-like coproduct for the state $\rho$, provided that the 
correction terms can be written as perfect squares. These possibilities are explored in more detail for one case in the appendix, \S \ref{sec:FurtherPoss}. There is it shown that an (outer) ARW based on such an extended, non-group-like state, now has a `polypodic' aspect in that there are \emph{two} sets of height coordinates for each location $n \in {\mathbb N}$,
$\{d_n \}_{n\in {\mathbb N}}$ and $\{c_n\}_{n\in {\mathbb N}}$, which get augmented with different probabilities, under evolution by mixtures of analogously parametrised series. We do not pursue these more elaborate possibilites further, but continue with the examination of the role of different products and coproducts in the symmetric function context.

%%%%%%%%%%%%%%%%%%%%%%%%%%%%%%%%%%%%%%%%%%
\subsection{Inner ARW for $(\Lambda, \delta)$}
\label{subsec:InnerARW} 
We now turn to the investigation of ARWs for the second available product carried in the ring of symmetric functions, namely the \emph{inner} symmetric function product. The principles from the general setting will be carried through with adjustments now made for inner products. We present here two cases, firstly a \emph{pure} inner ARW, which is quite different from the outer ARW, as we will see; secondly, we consider a \emph{mixed} inner- and outer- ARW which can be seen as an extension of the outer case.

Proceeding as before, we define the pure inner ARW walk transformation,
\[ 
T{}^{in}_\psi(f) = \sum  \langle \psi \mid f_{[1]}\rangle f_{[2]}
\]
now invoking the \emph{inner} coproduct for symmetric functions $f$, and anticipating that $\psi$ is also a symmetric function. 
For the (initial) density $\rho $ we have $\rho(f) = \langle \rho \mid  f\rangle $. 

Note that the unit of \emph{inner} multiplication is not the scalar unit $1_{m}= \eta(1) :=s_{(0)}(X)$, but rather is now a symmetric function series, this time $1_{\sf m} \equiv  M$ playing a different role (note the expansion 
in complete symmetric functions and power sums, $M= \sum_n h_n= \sum_\lambda p_\lambda/z_\lambda$). Correspondingly, the co-unit $\varepsilon^{\sf m}$ is now\footnote{
Indeed, the Schur-Hall scalar product can itself be written in terms of the inner product, $\langle f \mid g \rangle = \varepsilon^{\sf m}(f * g)$, as can be established in the power sum basis where $p_\lambda * p_\mu = z_\lambda p_\lambda \delta_{\lambda,\mu}$, and $\langle  M \mid p_\lambda \rangle=1$.} evaluation with the scalar product with respect to $M$, $\varepsilon^{\sf m} = \langle M \mid \cdot ~ \rangle$. If normalisation is taken with respect to this unit we have
\begin{align}
1= \rho(1_{\sf m}). \nonumber
\end{align}
Similarly, positivity is demanded for $\rho$ measuring symmetric functions of the form $f\ip f$, requiring 
\[
0 \le \langle \rho \mid f\ip f \rangle = \sum  \langle \rho_{[1]} \mid f\rangle \, \langle \rho_{[2]}\mid f\rangle
\]
using the compatibility between inner product and inner coproduct.
However, group-like elements for the inner coproduct are simply the power sum functions $p_\lambda$. We proceed
by considering then a general expansion $\rho = \sum r_\lambda  p_\lambda$ in the power sum basis for arbitrary coefficients $r_\lambda$, and similarly for a test function, $f = \sum  f_\mu p_\mu$. The normalisation condition reads explicitly,
\begin{align}
1 = & \, \rho(1_{\sf m}) = \langle M \mid \rho \rangle =  \sum_\lambda r_\lambda \langle M \mid p_\lambda \rangle  = \sum_{\lambda} r_\lambda, \nonumber
\end{align}
so that the state $\rho$ is indeed normalised provided it is a convex combination of power sums. Turning to the condition for positivity, we have explicitly
\begin{align}
f*f = & \, \sum_{\lambda,\mu} f_\lambda f_\mu \, p_\lambda*p_\mu = \sum_\lambda f_\lambda^2 z_\lambda p_\lambda ,\nonumber \\
\langle \rho \mid f*f \rangle = & \, \sum_{\lambda,\mu} r_\mu f_\lambda^2 z_\lambda \langle p_\mu \mid p_\lambda \rangle = \sum_\lambda  r_\lambda (f_\lambda z_\lambda)^2, \nonumber
\end{align}
as expected. 

Reading off the induced action on the state $\rho$ induced by the generator $\psi$, we have 
\[
T{}^{in}_\psi{}^*(\rho)\big(f\big) =  \langle \rho \mid T{}^{in}_\psi(f)\rangle = \sum \langle \psi \mid f_{[1]}\rangle \langle\rho \mid  f_{[2]}\rangle = \langle \psi \ip \rho\mid f\rangle = \psi*\rho\,(f)
\]
using again the inner product-coproduct duality. Analogously to the previous case, the induced action on the state (now the inner convolution, $\psi{\scriptstyle{{[}}}\!\!\!\!\star\!\!\!\! {\scriptstyle{{]}}}\rho$) is given by the relevant multiplication of symmetric functions, in this case inner multiplication rather than outer (pointwise) multiplication. In the power sum basis with $\psi = \sum \psi_\alpha p_\alpha$, $\rho = \sum_\beta r_\beta p_\beta$, and 
$f = \sum_\gamma f_\gamma p_\gamma$, say, this gives explicitly with $\rho(f)  =  \sum_\lambda r_\lambda (z_\lambda f_\lambda)$,
\begin{align}
\psi \ip \rho = & \, \sum_\lambda (\psi_\lambda r_\lambda z_\lambda) p_\lambda, 
&& \psi \ip \rho(f) =  \sum_\lambda (\psi_\lambda r_\lambda z_\lambda) z_\lambda f_\lambda;\quad  \nonumber \\
\delta(f) = & \, \sum_\lambda f_\lambda p_\lambda \otimes p_\lambda, 
  && \sum \langle \psi \mid f_{[1]}\rangle \langle\rho \mid  f_{[2]} \rangle = 
\sum_\lambda f_\lambda (\psi_\lambda z_\lambda)(r_\lambda z_\lambda).\quad \nonumber
\end{align}

Finally checking normalisation of the evolved state, with $\delta(M) = \sum_\lambda z_\lambda^{-1} p_\lambda \otimes p_\lambda$, 
\begin{align}
1 = &\, \psi*\rho(1_{\sf m}) = \langle M \mid \psi*\rho\rangle = \sum_\lambda (\psi_\lambda r_\lambda z_\lambda),
\nonumber \\
\sum \langle M_{{[}1{]}} \mid \psi \rangle \langle M_{{[}2{]}} \mid \rho \rangle 
= & \, \sum_\lambda  z_\lambda^{-1} \langle   p_\lambda \mid \psi  \rangle \langle   p_\lambda \mid \rho  \rangle
= \sum_\lambda  z_\lambda^{-1} (z_\lambda \psi_\lambda) (z_\lambda r_\lambda ).
 \nonumber
\end{align}
In contrast to the outer ARW, the apparent lack of manifest normalisation under an inner ARW evolution is a consequence of the fact that the inner multiplicative unit is not group-like. Specific parametrisations of families of evolution operators $\psi$ can be envisaged, generating a stochastic process via \emph{inner} convolutions acting on an initial density $\rho$, giving $\psi{\scriptstyle{{[}}}\!\!\!\!\star\!\!\!\! {\scriptstyle{{]}}}\rho$, $\psi {\scriptstyle{{[}}}\!\!\!\!\star\!\!\!\! {\scriptstyle{{]}}}\psi{\scriptstyle{{[}}}\!\!\!\!\star\!\!\!\! {\scriptstyle{{]}}}\rho$, $\psi {\scriptstyle{{[}}}\!\!\!\!\star\!\!\!\!{\scriptstyle{{]}}} \psi{\scriptstyle{{[}}}\!\!\!\!\star\!\!\!\!{\scriptstyle{{]}}}\psi{\scriptstyle{{[}}}\!\!\!\!\star\!\!\!\! {\scriptstyle{{]}}}\rho$,$\cdots$ (that is, $\psi * \rho$, $\psi * \psi*\rho$, $\cdots$) as normalised states. However, we shall not explore such possibilities in detail here. We summarise the structure of the pure inner ARW as follows:

\begin{quotation}
\emph{Pure inner-ARW on symmetric functions}\\
A generic state $\rho = \sum_\lambda r_\lambda p_\lambda$ normalised with $\sum r_\lambda =1$ evolves via a generator $\psi$ 
under inner ARW evolution to $\psi*\rho = \sum_\lambda \psi_\lambda r_\lambda z_\lambda p_\lambda$. Positivity is guaranteed for positive coordinates $r_\lambda, \psi_\lambda$ but normalisation of successive convolutions $\psi{\scriptstyle{{[}}}\!\!\!\!\star\!\!\!\! {\scriptstyle{{]}}}\rho$, $\psi {\scriptstyle{{[}}}\!\!\!\!\star\!\!\!\! {\scriptstyle{{]}}}\psi{\scriptstyle{{[}}}\!\!\!\!\star\!\!\!\! {\scriptstyle{{]}}}\rho$, $\cdots$ (that is, $\psi * \rho$, $\psi * \psi*\rho$, $\cdots$) is not, and requires special choices of parametrisations for suitable families of states and generators.
\end{quotation}

We leave this scenario and move to an alternative where the inner-ARW walker action is still defined with respect to the inner coproduct and convolution, but the normalization and positivity conditions are retained from the previous outer product walk. This suggests that walks of both types can coexist, provided the parametrisation is sufficient to give a closed action. 
The resolution is to note that $\exp(\sum_n c_n p_n/n) = \sum_\lambda c_\lambda p_\lambda/z_\lambda$ where again $c_\alpha = \prod c_{\alpha_i}$, and thus that the hitherto arbitrary coefficients $r_\lambda$ of $\rho$ which we have used up to now in the power sum basis should be appropriately restricted. Up to scaling by the factor $z_\lambda$, these $r_\lambda$ themselves become (multiplicative) power sums of the auxiliary alphabet $U$ which we discussed above.  Normalisation (with respect to $1_{m}$), and positivity (measured on symmetric functions of the form $f\cdot f$, the \emph{outer} product), is now guaranteed as before. It only remains to calculate the inner product action $\psi \ip \rho$ corresponding to the inner product walk:
\begin{align}
\psi \ip \rho = & \, \big( \sum_\lambda \psi_\lambda p_\lambda/z_\lambda \big) \ip 
\big( \sum_\mu c_\mu p_\mu /z_\mu \big) = \sum_\lambda \psi_\lambda c_\lambda p_\lambda/z_\lambda
\nonumber \\
\equiv  & \, \exp \big(\sum_n (\psi_n c_n) p_n/n \big).
\nonumber
\end{align}
The situation of having mixed walks is then summarised as follows. 
\begin{quotation}
\emph{Combined outer-ARW and inner-ARW on symmetric functions}\\
In the notation $M(X,\{c_n\}_{n\in {\mathbb N}}) = M_c$ used above, consider convex combinations of both outer-ARW walk steps (with parameters $\lambda_i$, $\{\varphi_i^n \}_n$, $\sum_i \lambda_i = 1$) and inner-ARW walk steps (with parameters $\mu_j$, $\{\psi_j^n \}_n$, $\sum_j \mu_j = 1$).  Under the combined action the state evolves as
\begin{align}
M_c(X) \rightarrow & \, \sum_i \lambda_i M_{c+\varphi^i}  \rightarrow \sum_{i,j}  \lambda_i \mu_j M_{\psi^j(c\!+\!\varphi^i)},
\nonumber \\
\mbox{or} \qquad \qquad
M_c(X) \rightarrow & \, \sum_j  \mu_j M_{\psi^j c}  \rightarrow \sum_{i,j}  \lambda_i \mu_j M_{(\psi^j c)\!+\! \varphi^i} \nonumber
\end{align}
with the following interpretation. Walker states at locations $\{n\}_{n\in {
\mathbb N}}$ are labelled with height coordinates $\{c_n\}_{\mathbb N}$. Under the evolution of the inner and outer product ARWs, these cordinates can be either multiplicatively scaled (inner) or additively augmented (outer), with appropriate probabilities. 
\end{quotation}

Remarkably, the primitive $\lambda$-ring operations $X \rightarrow X+Y$, $X\rightarrow XY$ at the level of the underlying alphabet of the symmetric function ring, happen to be reflected in the \emph{arithmetic} operations of addition $+$ and multiplication $\cdot$ on the coordinates of the distribution $\rho$ (see also \cite{fauser:jarvis:2006a}, appendix).
%%%%%%%%%%%%%%%%%%%%%%%%%%%%%%%%%%%%%%%%%%
\subsection{Plethystic ARW for $(\Lambda, \nabla)$}
\label{subsec:PlethysticARW} 
\mbox{}
The third type of symmetric function ARW proceeds from the basic formalism via the plethysm coproduct $\nabla$. Recall that this is dual to symmetric function composition. This operation is noncocommutative but coassociative, and by definition compatible with the usual scalar product, but is not matched in a natural bialgebra with any product in $\Lambda$. However, this structure is enough to define the plethystic or exponential random walk which we investigate here. 

Note that we can define both a left- and a right- plethystic ARW depending on which of the Sweedler parts of
the convolution of the target function is acted on in the walker step:
\[
{}^LT^{pleth}{}_\phi(f) = \sum (\phi, f_{\langle 1\rangle}) f_{\langle 2\rangle},
\qquad
{}^RT^{pleth}{}_\phi(f) = \sum (\phi, f_{\langle 2\rangle}) f_{\langle 1\rangle}.
\]
The difference between walks is emphasized by the induced actions, namely 
${}^LT^{pleth}_\phi(\rho) = \phi[\rho]$, ${}^RT^{pleth}_\phi(\rho) = \rho[\phi]$, respectively.

The general evaluation of the walker steps is a difficult computational 
task in general. For algebraic simplicity, and to enable the analysis to be completed for the purposes of 
illustration, and comparison with the other classes of walks, we concentrate here on a special case of the \emph{right} plethystic ARW (and drop the ${}^R$ superscript hereafter). Below we briefly discuss the general case, and also compare and contrast the right- and left- cases but without explicit calculation.

%For definiteness we take $\phi = \exp(p_m)$, that is, we work with the special case $\varphi_k = m \delta_{km}$.
As before we are working with the standard set of normalisation and positivity conditions, and so the parametrisations of the walker actions and of the distribution function must be compatible. 
Thus we must evaluate $\rho[\phi]$ where $\rho = M_c = \exp(\sum c_np_n/n)$. We take the special case $\phi = p_m$, a \emph{single} elementary power sum, and then take up generalisations.

Various rules govern the evaluation of symmetric function plethysms \cite{littlewood:1940a}. The the salient ones are distributivity (generalised Laplace pairing) rules of addition from the left and right, $(A+B)[C] = A[C]+B[C]$, $A[B+C] = A_{(1)}[B] A_{(2)}[C]$, as well as that of  of outer multiplication on the left and right, $AB[C] = A[C]B[C]$, $A[BC] = A_{[1]}[B] A_{[2]}[C]$. These lead to the following
\begin{quotation}
\emph{Right-plethystic ARW generated by $p_m$}\\
Under ${}^R T_\phi^p$ for $\phi$ a single power sum $p_m$ we have
\begin{align}
M_c{[}p_m{]} =  & \, 
\exp \big( \sum_n c_{n} p_n[p_m]/m \big) \equiv  \exp \big( \sum_n (m c_{n}) p_{mn}/(mn) \big),  \nonumber
\end{align}
with the following tentative interpretation.
There occurs a combinatorial inflation of the coordinates $n$ of the height distribution $\{c_n\}_n$  of the state $M_c$. Locations $n=0 \mod m$ acquire heights $m c_{n/m}$,  while heights at locations $n \ne 0 \mod m$ are set to zero.
\end{quotation} 

For a weighted combination 
$\phi = w_1 p_{m_1} + w_2 p_{m_2} + \cdots + w_{\scriptsize{K}} p_K$ however, we require the inner coproduct rule. Let $w = \prod_i w_i$ and let $\omega$ be the list of $\{m_i\}_{i=1}^K$ arranged as a partition (so that the product of the power sums is
$p_\omega = \prod_i p_{m_i}$). Then since $\delta^{(K\!-\!1)} p_n = \otimes^K p_n$ and each scaled power sum $w_i p_p$ can be interpreted as deriving from a multiply copied underlying alphabet (formally extended from whole number to arbitrary scalar multiples), we have
\begin{align}
p_n[\sum_i w_i p_{m_i}] = & \, \prod_i p_n[w_i p_{m_i}] 
\equiv  w  p_{n\omega}, \nonumber \\
\mbox{so} \qquad
\exp\big(\sum c_np_n/n\big)[\phi] = & \, \exp\big(\sum ( wc_n) p_{n\omega}/n\big). \nonumber
\end{align}
The situation here is similar to that of the generalised outer ARW walk steps considered using non-group-like series for $\phi$. In order to accommodate this more general type of combinatorial inflation, the state $M_c$ would require coefficients giving `heights' for \emph{arbitrary} partitions $\mu$ in the exponent, rather than just the elementary power sums $p_n$ with coefficients $c_n$. Thus, as it stands, the parametrisation is not closed
under such more general plethystic steps (but see appendix, \S \ref{sec:FurtherPoss}, for a discussion of more general series parametrisations for which closure under such moves would be possible).
%%%%%%%%%%%%%%%%%%%%%%%%%%%%%%%%%%%%%%%%%%
%%%%%%%%%%%%%%%%%%%%%%%%%%%%%%%%%%%%%%%%%%
%%%%%%%%%%%%%%%%%%%%%%%%%%%%%%%%%%%%%%%%%%
%%%%%%%%%%%%%%%%%%%%%%%%%%%%%%%%%%%%%%%%%%
\section{Conclusions}
\label{sec:Conclusions}

In this work we have considered the ARW formalism in the context of the ring of symmetric functions, as equipped with various products, coproducts and bialgebraic structures. The interplay of these has enabled a generalised  type of random walk to be identified, based on parametrised states derived from certain formal infinite series, well-known in the theory of symmetric functions (as reviewed in \S \ref{sec:SymmFns} above). Series with group-like coproduct are naturally singled out to ensure positivity of measures in this real context, however some additional non-group-like series were also identified as candidate states satisfying positivity. Analogously to the classical random walk on the line, there are discrete walker locations (in this case ${\mathbb N}$), and walk steps are implemented by rearranging `height' coordinates representing the walker occupancy of the locations. Different symmetric function coproducts involved in the convolutions operative in the evolutionary steps lead to different modifications. It was found that local translations (additive changes to heights) and local scalings (multiplicative changes to heights) are generated by symmetric function outer and inner coproducts, respectively. The (restricted) type of plethysm coproduct  considered led instead to global inflations of the sequence of height coordinates (for example, decimation, whereby heights are reallocated at intervals of say 10 units, leaving intervening zeroes). These results are summarised in \S\S \ref{subsec:OuterARW}, \ref{subsec:InnerARW}, and \ref{subsec:PlethysticARW} above.

In the light of our study is worth emphasizing again the close structural similarity, noted in the introduction, between aspects of the symmetric function ARWs and one of the original case studies, the transcription of the random walk on the line into the ARW notation (in the algebraic language, an ARW on $Fun({\mathbb R})$ viewed as a Hopf algebra \cite{majid:1993qrwtr}). The role of the point measure $\delta(x-a)$ in allocating walker locations (for which of course $\int f(x)\delta(x-a) dx = f(a)$) is closely mirrored by that of the Cauchy kernel $M(X,U)$ in the symmetric function case, and the integral over ${\mathbb R}$ in evaluating averages of course corresponds to the Hall-Littlewood scalar product $\langle \cdot \mid \cdot \rangle_\Lambda$ for which we have the reproducing property, for any symmetric function,
$\langle M(X,U) \mid f(X) \rangle_{\Lambda(X)} = f(U)$ as we have seen. In this vein it should be noted that variations on the theme of symmetric functions, with modifications to the scalar product, $q$-symmetric functions, and non-commutative versions, are manifold and could equally provide further examples of interesting ARWs.

Consideration of asymptotic aspects or continuum and classical limits of the symmetric function walks are deferred to further work and beyond the scope of this paper. As far as the iterated ARW evolution is concerned we make the following observation related to character theory. The map (\ref{eq:GLbranch}) above can be seen in terms of group characters as the branching rule whereby a $GL$-character decomposes into a sum of products of subgroup $GL$-characters under restriction. For finite rank this is the standard identification of $GL(m)\times GL(n)$ as a subgroup of $GL(m+n)$; in the inductive limit the mapping and its iteraction are to be interpreted as 
$GL(\infty) \supset GL(\infty)\times GL(\infty) \supset GL(\infty)\times GL(\infty)\times GL(\infty) \supset \cdots$, thus the effect of the walk on random variables (Schur functions) is that of an infinite cascade of $GL$-branchings.

The present work is related to investigations of Markov processes on partitions and the symmetric group \cite{vershik:okounkov:2004a,olshanski:borodin:2006mpp}. There the standard complex $z$-measure $M_z(\lambda)$ on random partitions is related to a %$\varepsilon^z(s_\lambda)$ specialization, whereas in the present work $\langle M(X,U) | s_\lambda(X) \rangle_{\Lambda}$ is 
specialization of $s_\lambda$ based on $X = (1^n, 0,0, \cdots)$ with $n$ continued to arbitrary $z$, whereas in the present work $\langle M(X,U) | s_\lambda(X) \rangle_{\Lambda}$ is precisely $s_\lambda(U)$ for the alphabet $U$ associated to the parameters $\{c_n\}$ by specialization. As is well known, the structure of the ring $\Lambda$ is intimately related to operator constructs such as the semi-infinite wedge product spaces used in the fermion-boson correspondence, and underlying the study of integrable models (see \cite{schmidt:schnack:2002a} for an introduction to symmetric functions and partition functions in statistical mechanics). Indeed variations on the series $M(X,t)$ (and their duals $M^\perp(X,t)$) provide the material for the vertex operators appearing in many constructions of representations of infinite-dimensional algebras, and their matrix elements have been well-studied in that context
(see also \cite{fauser:jarvis:king:2010a} where vertex operators for universal characters of the orthogonal and symplectic classical subgroups of $GL$, together with some non-classical analogues, are presented). The present study can be seen as facilitating a link between these topics and the ARW programme. 
\vfill
\pagebreak
{\small

%\bibliography{qrw,sql}
%\bibliographystyle{plain}
%\def\topsep{0pt}
%\def\parsep{0pt plus 5pt minus 1pt}
%\def\itemsep{-0.5ex}
\vfill
\begin{appendix}
\section{Further possibilities for the outer ARW using symmetric function series}
\label{sec:FurtherPoss}
It is possible to relax the condition of having a group-like coproduct for the state $\rho$, provided that the 
correction terms can be written as perfect squares. Alternatively, the condition for positivity of ${T{}^{out}_\phi}^*(\rho)$ can be relaxed not only by allowing $\phi$ to be a convex sum of group-like elements, as above, but again by ensuring that $\phi$ is group-like up to corrections which are perfect squares.
Within the symmetric function series \cite{littlewood:1940a}, we consider as an illustration, the following extension of $M$:  
\[
M{[} s_1 s_1{]}(X,V) \equiv  M(XX,VV) =  \prod_{i,j,k,\ell} (1-x_ix_jv_kv_\ell)^{-1}
=\exp\big(\left.\sum\right._{k\geq1} p_{(k,k)}(X) p_{(k,k)}(V)/k \big)
\]
where the (inner) group-like property of the power sums is invoked -- for a general composite alphabet, we have $p_n(XY) =
p_n(X)p_n(Y)$, in this case applied to $Y \equiv X$, and similarly for $p_n(VW)$ for $W\equiv V$.  We postulate a corresponding generalised state $M(XX,VV) M(X,U)$, say ${M'}{}(X,U,V)$ or, symbolically, $M_{d,c}$ involving \emph{two} auxiliary alphabets $U,V$ with specializations $p_k(U) \leftrightarrow c_k$, as before, but now $p_{(k,k)}(V) = p_k(VV) = p_k(V){}^2 =d_k^2$. The reason for doubling the second alphabet $V$ becomes clear when the outer coproduct is examined\footnote{Extending the original alphabet of indeterminates $X$ to $X,Y$, exploiting $\Lambda(X,Y) \cong \Lambda(X)\otimes \Lambda(X)$.}:
\begin{align}
{M'}{}(X,Y,U,V) = & \, M(XX,VV) M(X,U) M(YY,VV) M(Y,U) M(XY,VV) \nonumber \\
= & \, {M'}{}(X,U,V) {M'}{}(Y,U,V)\cdot M(XY,VV) \nonumber \\
= & \, \sum_\alpha  {M'}{}(X,U,V) {M'}{}(Y,U,V) s_\alpha(XV) s_\alpha(YV),
\nonumber
\end{align}
where the last form comes from the Cauchy identity for $M(XV,YV)\equiv M(XY,VV)$ by recognising the free product over all indeterminates $x_iy_jv_kv_\ell$ as quadratic in the composite alphabets $XV$ and $YV$.
The corresponding evaluation on $f^2$ (once again using the Laplace property) is now a sum of squares:
\[
\langle {M'}{}(X,U,V) \mid f^2 \rangle_{\Lambda(X)} = \sum_\alpha
  \langle {M'}{}(X,U,V)s_\alpha(XV) \mid f \rangle_{\Lambda(X)}^2
\]
  
Schematically, the ARW based on this extended, non-group-like state, now has a `polypodic' aspect in that there are two sets of height coordinates for each location $n \in {\mathbb N}$,
$\{d_n \}$ and $\{c_n\}$, which get augmented with different probabilities, under evolution by mixtures of analogously parametrised series. We do not pursue these more elaborate possibilites further in the present work\footnote{Note the possibility of exploiting the presence of $p_k(XX) = p_{k,k}(X)$ terms in the state parametrisation, as proposed here for $M_{d,c}$, in order to accommodate more general moves in the plethystic type ARW's (see \S \ref{subsec:PlethysticARW} above).}.
\vfill
\end{appendix}
\pagebreak
{\small\tableofcontents}
\vfill
\printindex
}

\end{document}